\documentclass{emulateapj}

\usepackage{epsfig} 

\newcommand{\beq}{\begin{equation}}
\newcommand{\eeq}{\end{equation}}
\newcommand{\beqa}{\begin{eqnarray}}
\newcommand{\eeqa}{\end{eqnarray}}

\def\fun#1#2{\lower3.6pt\vbox{\baselineskip0pt\lineskip.9pt
  \ialign{$\mathsurround=0pt#1\hfil##\hfil$\crcr#2\crcr\sim\crcr}}}

\shorttitle{Reconstructing and Deconstructing Dark Energy}
\shortauthors{Linder}

\begin{document} 

\title{Reconstructing and Deconstructing Dark Energy} 
\author{Eric V. Linder} 
\affil{Physics Division, Lawrence Berkeley Laboratory,
Berkeley, CA 94720} 
\email{evlinder@lbl.gov}

\begin{abstract} 
The acceleration of the expansion of the universe, ascribed to a 
dark energy, is one of the most intriguing discoveries in science. 
In addition to precise, systematics controlled data, clear, robust 
interpretation of the observations is required to reveal the nature 
of dark energy.  Even for the simplest question: is the 
data consistent with the cosmological constant?\ there are important 
subtleties in the reconstruction of the dark energy properties.  We 
discuss the roles of analysis both in terms of the Hubble expansion 
rate or dark energy density $\rho_{DE}(z)$ and in terms of the dark 
energy equation of state $w(z)$, arguing that each has its carefully 
defined place.  Fitting the density is best for learning about the density, 
but using it to probe the equation of 
state can lead to instability and bias. 

\end{abstract} 



\section{Introduction} \label{sec.intro}

The acceleration of the expansion of the universe represents a 
challenge to our understanding of fundamental physics.  Whether the 
resolution of this mystery lies in high energy physics, the theory of 
gravitation, the nature of the quantum vacuum, extra dimensions, etc.\ 
remains unknown.  The simplest model, and possibly one motivated from 
string theory \cite{kachru}, is Einstein's cosmological constant, though 
it too is fraught with complications, e.g.\ the fine tuning and coincidence 
problems \cite{weinberg,carroll}. 

Nevertheless, the cosmological constant serves as a benchmark, a first 
indication for how drastically we might need to extend our physics 
framework to 
understand the acceleration of the universe.  Additionally, because this 
model offers such definite and unvarying predictions for the dark energy 
properties, it gives a robust target.  In this paper we consider various 
aspects of interpretation of the data, present and future, that aid or 
obscure the goal of determining the nature of dark energy and its 
consistency with the cosmological constant.  In the sense that conflicting 
or biased interpretations of the data, rather than the data itself, affect 
what we determine to be the reality, we refer to this problem as 
``deconstruction'' of dark energy. 

\section{Reconstruction and Deconstruction} \label{sec.recon}

The cosmological constant has an energy density that is constant 
in time, $\rho(z)=\rho(0)$ in terms of redshift $z$, a 
pressure the negative of the energy density, $p=-\rho$, or 
equivalently an equation of state ratio (EOS) $w\equiv p/\rho=-1$ constant 
in time, and no 
spatial fluctuations in $\rho$ or $w$.  So we could think of testing 
whether the dark energy is a cosmological constant by seeing if the 
observations are consistent with either a constant density or $w=-1$. 

\subsection{Constant equation of state $w_{\rm const}=-1$} 

The first point to make is that fitting the data in terms of a 
constant EOS parameter is insufficient.  Many models with time variation 
$w'$ can be fit, for data of limited precision or over a limited redshift 
range, by a $w_{\rm const}$.  But this obscures the physics behind the 
dark energy.  Even if we are so modest and undemanding as to only ask 
if $w_{\rm const}=-1$ or not, we can be fooled by time varying models 
mimicking this value. 

\begin{figure}[!hbt]
\begin{center} 
\psfig{file=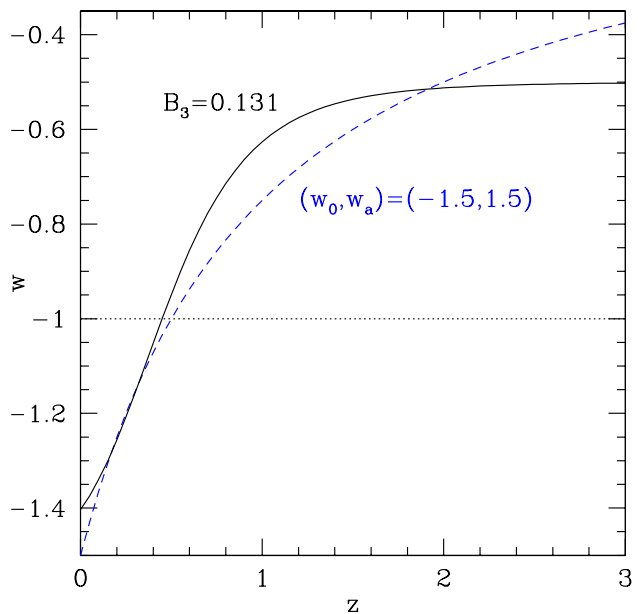,width=3.4in} 
\caption{Dark energy models possessing strong time variation can still 
appear consistent with cosmological constant when interpreted only in 
terms of a constant equation of state.  The two models shown here are 
consistent with all current data and a $w_{\rm const}=-1$ cosmological 
constant model. 
} 
\label{fig.wconst}
\end{center} 
\end{figure}

We illustrate this in Figure \ref{fig.wconst}.  The two dark energy 
models shown, one a high energy physics model parametrized with a 
time evolution $w(a)=w_0+w_a(1-a)$, where $a=1/(1+z)$ is the scale 
factor of the universe, with $(w_0,w_a)=(1.5,-1.5)$, and one an 
extended gravity model (case 3 of \citealp{lingrav}) 
modifying 
the Friedmann expansion equation, both possess a 
strong time variation in the equation of state.  However, in comparison 
to present data both appear consistent with a cosmological constant 
when interpreted in terms of a $w_{\rm const}$.

Both models are consistent, to 0.1\%, with the distance to the 
last scattering surface of the cosmic microwave background (CMB) in 
the cosmological constant model with the same matter density.  Even 
the next generation Planck experiment \cite{planck} will not be able to 
distinguish these models from the cosmological constant on this basis. 
With regard to large scale structure, the linear theory growth factors 
agree to better than 4\%.  This roughly corresponds to requiring 
knowledge of the mass fluctuation amplitude $\sigma_8$ at this level, 
but the current systematic uncertainties appear to be of order 20\%.  Most 
constraining are the Type Ia supernova distances, for which the time 
varying models deviate from the cosmological constant by under 0.1 
magnitude.  This is currently at the $1\sigma$ limit of precision. 

So if our universe followed either of these dark energy models (or 
a host of others that could be chosen) then by using the $w_{\rm const}$ 
paradigm we would happily interpret 
the data as pointing to a cosmological constant and miss the key 
clues to new physics. 

Of course as cosmological probes become more precise, we will be able 
to distinguish the illustrated models from the cosmological constant. 
However if the interpretation does not include the possibility of 
time variation in the dark energy equation of state we will again 
miss the physics and think the particular models shown are 
$w_{\rm const}=-1.2$ models.  And indeed dark energy models with 
a less extreme time variation than shown in Fig.\ \ref{fig.wconst} 
might well look like $w_{\rm const}=-1$ to within the data precision. 
Moreover, many models 
have an asymptotic or attractor behavior that brings the EOS very near the 
cosmological constant value for a period of the expansion history. 
Examples include the linear potential model \cite{linde,kallosh}, 
the cyclic model \cite{turok}, the ``ripstop'' model where particle 
production offsets the superacceleration of phantom energy with 
$w<-1$ \cite{lingrav}, and certain scalar-tensor theories \cite{matarrese}. 

For a window onto the new physics we 
need to look for both the present or averaged EOS value and a measure 
of its time variation, say $w'$.  Note that throughout this paper we 
refer to the EOS function $w(z)$ as an effective EOS, defined in terms 
of the expansion history $a(t)$, and not necessarily restricted to a 
physical component like a scalar field.  From \citealp{linjen} (cf.\ 
\citealp{staro}) we have 
the effective EOS defined by 
\beq
w(z)=-1+\frac{1}{3}\frac{d\ln \delta H^2}{d\ln (1+z)}, \label{eq.wdh}
\eeq 
where the Friedmann expansion equation reads 
\beq
H^2(z)/H_0^2=\Omega_M(1+z)^3+\delta H^2(z). 
\eeq 
The Hubble parameter $H=\dot a /a$ and $\Omega_M$ is the matter 
density, so $\delta H^2$ represents our ignorance of either components 
beyond matter or of extensions to general relativity that modify the 
Friedmann equation. 

A very robust and convenient measure of the time variation 
of the dark energy EOS is given by the two parameter form 
\citep{linprl,lingrav}
\beq 
w(z)=w_0+w_a(1-a)=w_0+w_az/(1+z). 
\eeq 
We see that this implies a characteristic variation 
\beq 
\partial_N w\equiv -\frac{dw}{d\ln a}=aw_a. 
\eeq 
Note that this is closely related to the inflationary theory measure 
of time variation: 
\beq 
\partial_N w=-dw/dN,
\eeq 
where $N=d\ln a$ counts the number of e-folds of expansion during inflation. 
Just as in inflation, there is ``running'' -- a variation of parameters 
during the acceleration -- and we need to choose when to evaluate the 
derivatives, e.g.\ at $N=60$.  The derivative of the EOS will 
also be a function of the redshift when it is evaluated, and for 
simplicity we define a number from the function: $w'\equiv\partial_N 
w(z=1)=w_a/2$.  
This is motivated by $z\approx1$ being approximately when dark energy 
begins to become dynamically important for the expansion. 

So we have argued that to learn the direction of new physics 
we need to consider not only the value of 
$w_{\rm const}$ but not deny the possibility of time variation $w'$.  
That is, if we want to 
say that observations point to the cosmological constant, we need to 
see that $w'\approx0$.  A much more modest goal is merely to ask 
whether the observations are consistent with the cosmological constant. 
In this case, we might consider looking for whether the dark energy 
density is constant, rather than addressing the physics question of its 
equation of state.  

\subsection{Constant density} \label{sec.22} 

Note that reconstructing a constant energy density is only a necessary, 
not sufficient, condition for a cosmological constant.  Some models 
might arrange an effective energy density nearly 
constant with time yet with effective pressure not equal to the 
negative of the energy density; this holds in some k-essence and tachyon 
field models (cf.\ \citealp{scherrer}).  So mapping the density alone contains 
insufficient information if we want to remain open 
to the possibility that the unknown physics could involve components 
other than canonical, minimally coupled scalar fields or involve 
extensions to gravity. 

But in asking for consistency, not insight, seeking to reconstruct 
the dark energy density behavior (see e.g.\ 
\citealp{wangtegmark,freese,hutcoo}) 
is a perfectly acceptable method.  Testing for the EOS $w(z)$ also allows 
one to check consistency (plus more of course), and one can also build 
up the density history through an integration: 
\beq 
\rho(z)=\rho(0)\,e^{3\int_0^z dz'\,[1+w(z')]/(1+z')}.
\eeq 
Of course this will diminish the sensitivity of the reconstruction of 
$\rho(z)$ somewhat, though not drastically \citep{hutcoo}.  
Also, it may be easier to test for a specific value, e.g.\ $w=-1$, 
than for a property, e.g.\ $\rho={\rm constant}$ \citep{hutpriv}.  
Generally, if one is only interested in whether the density is constant, 
use reconstruction in terms of the density (though in terms of EOS is 
not unreasonable).  However, if one is interested in the physics behind 
the dark energy, the EOS reconstruction is strongly preferred. 

This is because going from $\rho(z)$ to $w(z)$ requires differentiation, 
whether explicitly or implicitly, and this leads to instability and bias 
in the results.  This is what we referred to in the introduction as the 
deconstruction problem.  The dangers of explicit differentiation of the 
data are well known \citep{huttur,welal,teg02}, 
but those of implicit differentiation 
have been illustrated more recently.  \citealp{jonsson} pointed out that 
assuming a functional form for the density, or Hubble parameter, could 
distort conclusions regarding the equation of state, even apparently 
preferring a time evolving dark energy to fit data of a cosmological 
constant universe. 

Here we extend this to the general translation from a density reconstruction 
to an equation of state behavior.  Consider a set of values $\{\rho_i\}$, 
representing the density in redshift bins centered at $z_i$.  To obtain 
the EOS, one can imagine fitting the reconstructed $\rho_i$ with a 
polynomial or spline interpolation $\rho(z)$ (cf.\ \citealp{wangtegmark}) 
and then carrying out the derivative to form 
\beq 
w(z)=-1+\frac{1}{3\rho}(1+z)\frac{d\rho}{dz}. \label{eq.drho} 
\eeq 

In terms of the binned values, 
\beqa
w(z)=-1 &+& \frac{1+z}{3\rho_i}\Big\{\frac{\rho_{i+1}-\rho_i}{z_{i+1}-z_i} 
 \label{eq.spline} \\ 
&\quad& -\,\frac{z_{i+1}-z_i}{6} \rho_i''\left[\frac{3(z_{i+1}-z)^2}{(z_{i+1} 
-z_i)^2}-1\right]  \nonumber \\ 
&\quad& +\frac{z_{i+1}-z_i}{6} \rho_{i+1}'' 
\left[\frac{3(z-z_i)^2}{(z_{i+1}-z_i)^2}-1\right]\Big\}, \nonumber 
\eeqa 
where the second derivatives are determined in terms of the free parameters 
(bin values $\rho_i$) through the cubic spline formula and boundary 
conditions.  Thus we can write Eq.\ \ref{eq.spline} illustratively as 
\beq 
w(z)=-1+[1/(3\rho)](1+z)\sum c_i\rho_i, 
\eeq 
where $\rho_i$ are the free parameters of the problem and $c_i$ are 
weight functions.  One can imagine these being more general than 
cubic spline coefficients and instead being either polynomial expansions 
in redshift, e.g.\ $c_i=b_i(z-z_i)^n$, or involving orthogonal basis 
functions, so the following conclusions should be fairly general. 

Following the method of \citealp{jonsson}, now consider the stability of 
this deconstruction to small perturbations $\delta\rho_i$.  The quantity 
in braces in Eq.\ \ref{eq.spline} (or its generalizations) will involve 
terms like $\delta\rho_i$, $(\partial\rho_i''/\partial 
\rho_i)\delta\rho_i\sim\delta\rho_i$.  However, formally, because 
of the $1+z$ factor in front of the braces, a small perturbation to the 
fit to the data in terms of $\rho_i$ will run away at high redshift in 
terms of $w(z)$, showing the instability of this approach\footnote{It 
also may be problematic to take a linear interpolation 
as one of the spline boundary conditions as \citealp{wangtegmark} do.  
(Note that the $\rho'(z=0)=\rho(z_1)/z_1$ appearing there is a 
typo for $\rho'(z=0)=[\rho(z_1)-\rho(0)]/z_1$.)  This may allow 
statistical excursions of 
the density off the cosmological constant behavior to cause 
$w\ne-1$.}.  (Note that redshifts 
not much higher than 1 may be sufficient; \citealp{jonsson} find instability 
regions where $|w|>15$ for $z<1.8$.) 

Furthermore, this approach not only mediates against 
the cosmological constant (or any $w_{\rm const}$ model), but favors a 
model where the EOS crosses from less than $-1$ to greater than $-1$: 
precisely what is seen in current reconstructions.  This occurs because 
the quantity in braces is redshift dependent and will have a zero at 
some redshift $z_*$.  That is, for generic sets $\{\rho_i\}$ one will find 
$w(z_*)=-1$ (not necessarily within the redshift range of the data).  
Thus the deconstruction legislates for crossing the value $-1$. 
Combining these two effects together, a small perturbation $\delta\rho_i$ 
will generate a larger $\delta w$, forcing $w$ off from $-1$ (if the 
cosmological constant were the true fit) {\it and\/} forcing $w$ to 
cross $-1$ -- both leading to evolution. 

In summary, using $\rho_i$ fitting to find $\rho(z)$ is fine, but using 
$\rho_i$ to discuss $w(z)$ is perilous.  The latter generates errors 
of two types: false negatives where a true 
cosmological constant appears not to be so, and false positives where 
any model looks, over some finite redshift interval, not very distinct 
from a cosmological constant! 

That said, there appears to be a simple, if partial fix.  We can 
rewrite Eq.\ \ref{eq.drho} in terms of a logarithmic derivative with 
respect to scale factor $a$, rather than $1+z$.  If bins are defined 
in terms of $\rho(a)$, then the high redshift secular instability does not 
appear, and this should mitigate any tendency to cross the value $-1$. 
General issues of numerical differentiation, bin to bin, instability 
and the need for careful treatment of spline boundary conditions remain, 
but are possibly less severe. 

\subsection{Density or Equation of State?} 

What about the breadth of models encompassed by the two approaches? 
It has sometimes been claimed \cite{wangtegmark} that $w(z)$ suffers 
from not covering more esoteric possibilities like negative density. 
This is not wholly true, and even so is more of a feature than a bug. 
From Eq.\ \ref{eq.wdh} one sees that negative densities can be handled 
straightforwardly in terms of $w(z)$.  For example, consider a model 
where we (mistakenly) think $\Omega_M=1.2$ and find the best fit to 
data gives $\delta H^2=-0.2(1+z)^3$.  Clearly this looks like a negative 
density.  But $[(1+z)/\delta H^2]d\delta H^2/d(1+z)$ is well defined and 
gives $w(z)=0$: 
exactly as expected for this extra (negative) matter component. 

Where the $w(z)$ parametrization does blow up is not for negative 
density but for density crossing through zero.  In the parametrization 
of \citealp{wangtegmark}, they normalize the density by the present value, 
using $X(z)=\rho(z)/\rho(0)$, with of course $X(0)=1$.  In the negative 
density illustration above, $X$ will then always be positive.  It 
is only when $X$ goes from its defined value of 1 at $z=0$ to 0 at some 
redshift $z_*$ that $w(z_*)$ blows up to infinity.  But we can 
argue this is a feature not a bug.  

For example, in the linear potential 
model analyzed in \citealp{kallosh}, the dark energy density hits zero 
very shortly in cosmic time before the final ``cosmic doomsday'' 
collapse.  By parametrizing the $w(z)$ given by the linear potential 
(e.g.\ in terms of $w_0$ and $w_a$, contrary to the claim by 
\citealp{wangtegmark} that this model cannot treat the collapse case) 
and asking when it blows up (in the future), one obtains a very accurate 
estimate for the doomsday time.  

Furthermore, having $\rho$ cross through 
zero is not merely mildly esoteric, but violates either the Big Bang 
condition or the continuity equation 
\beq 
d\rho/d\ln(1+z)=3(\rho+p). 
\eeq 
Dividing through by $\rho(0)$, one has $dX/d\ln(1+z)=3(X+p/\rho(0))$. 
For acceleration, we take the pressure $p$ to be negative and so at 
the time when $X=0$, the left hand side will be negative and $\rho$ 
will be driven further negative at high redshift.  This negative 
energy density will cause the expansion rate $H$ to go to zero at some 
point in the past, describing a bounce universe rather than a Big Bang 
model.  If one insists on allowing the density to cross through zero 
then the price is either abandoning the continuity equation or the 
Big Bang early universe.  And if one abandons the continuity equation 
then the density alone is insufficient since pressure is unknown, 
leading back to an equation of state formulation. 

Finally, one can argue that the Friedmann equation of motion 
explicitly involves both energy density $\rho$ and pressure $p$, so a 
parametrization $w(z)=p(z)/\rho(z)$ is closer to the physics.  An 
example of this was pointed out at the beginning of \S\ref{sec.22}, 
where a constant energy density may not necessarily be matched by a constant 
negative pressure of equal magnitude.  Fundamentally, the passive 
gravitational energy $\rho+p$ and the active gravitational energy $\rho+3p$ 
both explicitly depend on more than merely the density, they involve an 
equation of state. 

\section{Conclusion} \label{sec.concl} 

Understanding, or at least obtaining insight into, the nature of dark 
energy will be the great challenge of physics in the next decade.  We 
must be sure that we ask questions in such a way that the answers we 
derive are not deconstructions -- subjective interpretations -- but 
faithful reconstructions of aspects of the true physics.  This 
includes allowing explicitly for the possibility of time variation in 
the equation of state of the dark energy.  Using a $w_{\rm const}$ 
will only tell us whether the data are consistent with the cosmological 
constant, not teach us that another, mimicking model is not the true 
answer.  

If our main concern is whether the dark energy density is constant in 
time, we can use reconstruction of the density values in (preferably 
uncorrelated) bins in expansion factor 
or a principal component analysis of the density (see 
\citealp{hutcoo,hutstark}).  Using the equation of state 
$w(z)$ in either bins or eigenmodes will give almost as accurate results, 
plus additional information.   Crosschecking one versus the other may 
also test exotic models involving violation of the continuity equation. 

In our quest to understand new physics, key clues are carried by 
the equation of state.  To reveal this one should fit for $w(z)$ directly. 
Using other, intermediary parametrizations such as density on the way 
to the equation of state can lead to 
unstable and biased solutions, both type 1 (false positive) and type 2 
(false negative) errors with respect to determining if the data are 
consistent with no time evolution, $w(z)=-1$, like for the 
cosmological constant model.  Of course systematics errors in the 
data themselves can also lead to false evolution.

\section*{Acknowledgments} 

This work has been supported 
in part by the Director, Office of Science, Department of Energy under 
grant DE-AC03-76SF00098.  I thank Alex Kim for useful comments, and 
especially the Michigan Center for Theoretical 
Physics for hospitality during the writing of this paper and Katie Freese, 
Paolo Gondolo, and Yun Wang for discussions there.

\end{document}